\documentclass{article}
\usepackage{graphicx}
\usepackage{amssymb,amsmath}
\usepackage{multirow}
\date{}

\makeatletter \@addtoreset{equation}{section}
\renewcommand{\theequation}{\thesection.\@arabic\c@equation}
\newcommand{\affiliation}[1]{\let\thefootnote\relax\footnote{\mbox{}\\ \noindent {#1}}}

\makeatother

\begin{document}
\title{\textbf{New Symbolic Algorithms For Solving A General Bordered Tridiagonal Linear System }}

\author{ A. A. KARAWIA\footnote{ Home Address: Mathematics Department, Faculty of Science, Mansoura University,
Mansoura 35516, Egypt. E-mail:abibka@mans.edu.eg}\\
Computer science unit, Deanship of educational services, Qassim University,\\
 P.O.Box 6595, Buraidah 51452, Saudi Arabia. \\
              E-mail: kraoieh@qu.edu.sa
}

\maketitle
\begin{abstract}
In this paper, the author present reliable symbolic algorithms for solving a general bordered tridiagonal linear system. The first algorithm is based on the LU decomposition of the coefficient matrix and the computational cost of it is $O(n)$. The second is based on The Sherman-Morrison-Woodbury formula. The algorithms are implementable to the Computer Algebra System (CAS) such as MAPLE, MATLAB and MATHEMATICA. Three examples are presented for the sake of illustration.\bigskip
\end{abstract}

\begin{flushleft}\footnotesize
\hspace*{0.9cm}{\textbf{Keywords}:Bordered tridiagonal matrices; LU factorization; Sherman-Morrison-Woodbury Formula; Computer\\
\hspace*{2.4cm} algebra systems(CAS). \\
\hspace*{0.9cm}{\textbf{AMS Subject Classification}:15A15; 15A23; 68W30; 11Y05; 33F10; F.2.1; G.1.0.\\}
}

\end{flushleft}
\newtheorem{alg}{Algorithm}[section]

\section{Introduction}

\hspace*{0.5cm} A general bordered tridiagonal linear system takes the form:
\begin{equation}
\left[
           \begin{array}{ccccccc}
             a_1 & b_1 & 0 & \cdots & \cdots& 0 & p_1 \\
             c_2 & a_2 & b_2 & \ddots & &\vdots & p_2 \\
             0 & c_3 & a_3 & b_3 &\ddots &\vdots & p_3\\
             \vdots & \ddots & \ddots & \ddots & \ddots& 0& \vdots \\
             \vdots &  & \ddots & \ddots & \ddots&\ddots & p_{n-2} \\
             0 & \cdots & \cdots &0 & c_{n-1} & a_{n-1}  & b_{n-1} \\
             q_1 & q_2 & \cdots& q_{n-3} & q_{n-2} & c_n & a_n \\
           \end{array}
         \right]\left[\begin{array}{c}
                  x_1 \\
                  x_2 \\
                  x_3 \\
                  \vdots \\
                  x_{n-2} \\
                  x_{n-1} \\
                  x_n
                \end{array}\right]=\left[\begin{array}{c}
                  y_1 \\
                  y_2 \\
                  y_3 \\
                  \vdots \\
                  y_{n-2} \\
                  y_{n-1} \\
                  y_n
                \end{array}\right]
,\quad n > 3.
\end{equation}
\\
Many problems in mathematics and applied science require the solution of a general bordered tridiagonal linear system. For example the solution of certain partial differential equations, spline approximation, computation of electric power system, etc. [1-4]. Recently in [5], the author presents an approach to find solution of the equation (1.1) in heterogeneous environments. In this approach, a bordered tridiagonal linear system is first converted into three or more tridiagonal linear systems that are independent each other, then the solution of the bordered tridiagonal linear system is obtained by solving the tridiagonal linear systems via parallel computing under heterogeneous environments. The motivation of the current paper is to establish efficient algorithms for solving a general bordered tridiagonal linear system of the form (1.1).\\

The paper is organized as follows: In Section 2, a new symbolic algorithm is constructed. The Sherman-Morrison-Woodbury algorithm is presented in Section 3. Three illustrative examples are given in Section 4. Conclusions of the work are presented in Section 5.

\section{A new Symbolic Algorithm}

\hspace*{0.5cm} In this section we shall focus on the construction of new symbolic algorithm for solving a general bordered tridiagonal linear system. To do this we
begin by considering the LU decomposition [6] of the coefficient matrix in (1.1)

\begin{equation}
\begin{split}
&\left[
           \begin{array}{ccccccc}
             a_1 & b_1 & 0 & \cdots & \cdots& 0 & p_1 \\
             c_2 & a_2 & b_2 & \ddots & &\vdots & p_2 \\
             0 & c_3 & a_3 & b_3 &\ddots &\vdots & p_3\\
             \vdots & \ddots & \ddots & \ddots & \ddots& 0& \vdots \\
             \vdots &  & \ddots & \ddots & \ddots&\ddots & p_{n-2} \\
             0 & \cdots & \cdots &0 & c_{n-1} & a_{n-1}  & b_{n-1} \\
             q_1 & q_2 & \cdots& q_{n-3} & q_{n-2} & c_n & a_n \\
           \end{array}
         \right]\\
         =&\left[
           \begin{array}{ccccccc}
             1 & 0 & 0 & \cdots & \cdots& 0 & 0 \\
             \frac{c_2}{d_1} & 1 & 0 & \ddots & &\vdots & 0 \\
             0 & \frac{c_3}{d_2} & 1 & 0 &\ddots &\vdots & 0\\
             \vdots & \ddots & \ddots & \ddots & \ddots& 0& \vdots \\
             \vdots &  & \ddots & \ddots & \ddots&\ddots & 0 \\
             0 & \cdots & \cdots &0 & \frac{c_{n-1}}{d_{n-2}} & 1  & 0 \\
             \alpha_1 & \alpha_2 & \cdots& \alpha_{n-3} & \alpha_{n-2} & \alpha_{n-1} & 1 \\
           \end{array}
         \right]\left[
           \begin{array}{ccccccc}
             d_1 & b_1 & 0 & \cdots & \cdots& 0 & \beta_1 \\
             0 & d_2 & b_2 & \ddots & &\vdots & \beta_2 \\
             0 & 0 & d_3 & b_3 &\ddots &\vdots & \beta_3\\
             \vdots & \ddots & \ddots & \ddots & \ddots& 0& \vdots \\
             \vdots &  & \ddots & \ddots & \ddots&\ddots & \beta_{n-2} \\
             0 & \cdots & \cdots &0 & 0 & d_{n-1}  & \beta_{n-1} \\
             0 & 0 & \cdots& 0 & 0 & 0 & d_n \\
           \end{array}
         \right]
\end{split}
\end{equation}

where $d_i$, $\alpha_i$ and $\beta_i$ in (2.1) satisfy:

\begin{equation}
d_i=\left\{\begin{matrix}
a_1  &\text{if}\quad i=1 \\
 a_i-\frac{b_{i-1}}{d_{i-1}}c_i &\quad\quad\quad\quad\quad\text{if}\quad i=2,3, ..., n-1 \\
 a_n-\sum_{j=1}^{n-1}\alpha_j\beta_j &\text{if}\quad i=n,
\end{matrix}\right.
\end{equation}

\begin{equation}
\alpha_i=\left\{\begin{matrix}
\frac{q_1}{d_1}  &\text{if}\quad i=1 \\
 \frac{1}{d_i}(q_i-\alpha_{i-1}b_{i-1}) &\quad\quad\quad\quad\quad\text{if}\quad i=2,3, ..., n-2 \\
 \frac{1}{d_{n-1}}(c_n-\alpha_{n-2}b_{n-2}) &\quad\text{if}\quad i=n-1,
\end{matrix}\right.
\end{equation}

and
\begin{equation}
\beta_i=\left\{\begin{matrix}
p_1  &\text{if}\quad i=1 \\
 p_i-\frac{\beta_{i-1}}{d_{i-1}}c_i &\quad\quad\quad\quad\quad\text{if}\quad i=2,3, ..., n-2 \\
 b_{n-1}-\frac{\beta_{n-2}}{d_{n-2}}c_{n-1}) &\quad\text{if}\quad i=n-1.
\end{matrix}\right.
\end{equation}
We also have the determinant of the coefficient matrix in (1.1):
\begin{equation}
\text{Determinant}=\prod_{i=1}^nd_i.
\end{equation}
At this point it is convenient to formulate our first result. It is a symbolic algorithm for
computing the solution of a general bordered tridiagonal linear system (1.1).

\begin{alg}
the solution of a general bordered tridiagonal linear system (1.1), we may proceed as follows: \\
\textbf{INPUT} number of equations in (1.1) $n$ and the components $a_i$, $i=1, 2, ..., n$, $b_i$, $i = 1, 2, . . . , n-1$, \\
\hspace*{1.5cm} $c_i$, $i = 2, 3, . . . , n$, $p_i$ and $q_i$, $i=1, 2, ..., n-2$.\\
\textbf{OUTPUT} The solution of a general bordered tridiagonal linear system $\mathbf{x}$.\\
\textbf{Step 1:} Set $d_1=a_1$, $\alpha_1=\frac{q_1}{d_1}$, $\beta_1=p_1$. If $d_1= 0$ then $d_1= t$($t$ is just a symbolic name) end if.\\
\textbf{Step 2:} For $i=2, 3, ..., n-1$\\
   \hspace*{2.5cm}Compute $d_i=a_i-b_{i-1}*c_i/d_{i-1}$,\\
   \hspace*{2.5cm}If $d_i= 0$ then $d_i= t$ end if.\\
\textbf{Step 3:} For $i=2, 3, ..., n-2$, compute\\
   \hspace*{2.5cm}$\alpha_i=(q_i-\alpha_{i-1}*b_{i-1})/d_i$,\\
   \hspace*{2.5cm}$\beta_i=p_i-\beta_{i-1}*c_i/d_{i-1}$.\\
\textbf{Step 4:} Set $\alpha_{n-1}=(c_n-\alpha_{n-2}*b_{n-2})/d_{n-1}$,\\
\hspace*{1.2cm} Set $\beta_{n-1}=b_{n-1}-\beta_{n-2}*c_{n-1}/d_{n-2}$,\\
\hspace*{1.2cm} Set $d_n=a_n-\sum_{j=1}^{n-1}\alpha_j*\beta_j$. If $d_n= 0$ then $d_n= t$ end if.\\
\textbf{Step 5:} Compute $\mathrm{Determinant} =\Big( \prod_{i=1}^nd_i\Big)_{t=0}$.\\
\textbf{Step 6:} Set $z_1=y_1$.\\
\textbf{Step 7:} For $i=2, 3, ..., n-1$, compute\\
\hspace*{2.5cm} $z_i=y_i-c_i*z_{i-1}/d_{i-1}$.\\
\textbf{Step 8:} Compute $z_n=y_n-\sum_{j=1}^{n-1}\alpha_i*z_i$, $x_n=z_n/d_n$, and $x_{n-1}=(z_{n-1}-\beta_{n-1} x_n)/d_{n-1}$.\\
\textbf{Step 9:} For $i=n-2, n-3, ..., 1$, compute\\
 \hspace*{2.5cm} $x_i=(z_i-b_ix_{i+1}-\beta_ix_n)/d_i$.\\
\textbf{Step 10:} Substitute $t=0$ in all expressions of the solution vector $\mathbf{x}$.\\
\end{alg}

The symbolic Algorithm 2.1 will be referred to as \textbf{SBTLS} algorithm. The computational cost of \textbf{SBTLS} algorithm is $19n-34$ operations. In [7], the \textbf{PERTRI} algorithm is special case of our algorithm when $p_i=q_i=0,\quad i=2,3, \cdots,n-2$.\\

\section{The Sherman-Morrison-Woodbury algorithm}
\hspace*{0.5cm} In this section, we are going to formulate a new symbolic algorithm for solving a general bordered tridiagonal linear system of the form (1.1) based on the  Sherman-Morrison-Woodbury formula and any symbolic tridigonal linear solver.\\

A general bordered tridiagonal linear system of the form (1.1) can be written in the form: \\
\begin{equation}
\left[\begin{array}{cc}
   M_1 & V \\
  U^T & M_2\\
  \end{array}
  \right]
  \left[\begin{array}{c}
   x^{'} \\
  x^{''}\\
  \end{array}
   \right]
   =\left[\begin{array}{c}
   y^{'} \\
  y^{''}\\
  \end{array}
   \right]
\end{equation}
where\\
$M_1=\left[
           \begin{array}{cccccc}
             a_1 & b_1 & 0 & \cdots & \cdots& 0 \\
             c_2 & a_2 & b_2 & \ddots & &\vdots  \\
             0 & c_3 & a_3 & b_3 &\ddots &\vdots \\
             \vdots & \ddots & \ddots & \ddots & \ddots& 0\\
             \vdots &  & \ddots & c_{n-2} & a_{n-2}& b_{n-2}  \\
             0 & \cdots & \cdots &0 & c_{n-1} & a_{n-1}   \\
           \end{array}
         \right], M_2=\left[
           \begin{array}{c}
           a_n\\
           \end{array}
         \right],
         V=\left[\begin{array}{ccccc}
             p_1 & p_2 & \cdots & p_{n-2} & b_{n-1}
           \end{array}\right]^T,\\
           U^T=\left[\begin{array}{ccccc}
             q_1 & q_2 & \cdots & q_{n-2} & c_n
           \end{array}\right],
           x^{'}=[x_1, x_2,\cdots,x_{n-1}]^T,\quad x^{''}=[x_n],\quad y^{'}=[y_1, y_2,\cdots,y_{n-1}]^T, \\\text{and} \quad y^{''}=[y_n].\\
         $

Thus (3.1) is equivalent to:\\

\begin{equation}\label{A_Label}
\begin{split}
M_1 x^{'}+Vx^{''}&=y^{'}\\
U^Tx^{'}+M_2x^{''}&=y^{''}
\end{split}
\end{equation}
Assume that $a_n\ne 0$. After elimination of $x^{''}$ from (3.2), we get the linear system:\\
\begin{equation}
Mx^{'}=y\hat{}
\end{equation}
where\\
\hspace*{1cm}$M=M_1-VM_2^{-1}U^T,\quad \text{and}\quad y\hat{}=y^{'}-VM_2^{-1}y^{''}.$\\
\\
If we applying the Sherman-Morrison-Woodbury formula[8] to M, we will obtain:\\

$M^{-1}=M_1^{-1}+M_1^{-1}V\left(M_2-U^T M_1^{-1}V\right)^{-1}U^TM_1^{-1},$\\
and\\

$x^{'}=M^{-1}y\hat{}=r+ M_1^{-1}V\left(M_2-U^TM_1^{-1}V\right)^{-1}U^Tr$.\\

where r is the solution of $M_1r=y\hat{}$. It is clear that the solution $x^{''}$ can be found from the above
formula by successive calculation of the expressions\\
$r=M_1^{-1}y\hat{},\quad q=M_1^{-1}V,\quad U^TM_1^{-1}V,\quad \left(M_2-U^TM_1^{-1}V\right)^{-1},\quad \text{and}\quad\left(M_2-U^TM_1^{-1}V\right)^{-1}U^Tr$.\\

The main part of above calculations is finding the first two expressions, which is equivalent to
solving two (n - 1)-by-(n - 1) tridiagonal linear systems with the same coefficient matrix $M_1$ and
different right-hand sides. After finding of $x^{'}$ , we can get $x^{''}$ from the second equation of (3.2) by formula\\
$$x^{''}=M_2^{-1}\left(y^{''}-U^Tx^{'}\right)$$\\

At this point it is convenient to formulate our second result. It is a symbolic algorithm for
computing the solution of a general bordered tridiagonal linear system of the form (1.1) and can be
considered it as a natural generalization of the The Sherman-Morrison algorithm in [7].\\

\begin{alg}
To compute the solution of a general bordered tridiagonal linear system (1.1), we may \\
\hspace*{2.7cm}proceed as follows:\\
\textbf{Step 1:} Find $M_1, M_2, U^T, V, y^{'}, y^{''},\text{and}\quad y\hat{}=y^{'}-VM_2^{-1}y^{''}$.\\
\textbf{Step 2:} Solve $M_1r=y\hat{}, \text{and} \quad M_1q=V$.\\
\textbf{Step 3:} Compute $x^{'}=r+ q\left(M_2-U^Tq\right)^{-1}U^Tr,\text{and}\quad x^{''}=M_2^{-1}\left(y^{''}-U^Tx^{'}\right)$.\\
\textbf{Step 4:} Compute the solution $x=\left[\begin{array}{c}
                                                x^{'} \\
                                                x^{''}
                                              \end{array}
\right]_{t=0}$\\
\end{alg}
Two systems $M_1r=y\hat{},\quad \text{and} \quad M_1q=V$ in algorithm 3.1 can be solved in parallel by any symbolic tridiagonal algorithm. The symbolic Algorithm 3.1 will be referred to as \textbf{SMWBTLS} algorithm.

\section{ILLUSTRATIVE EXAMPLES}
       In this section we give three examples for the sake of illustration.\\
\\
\textbf{Example 3.1.} Let\\
\hspace*{2.5cm}$
\left[\begin{array}{ccccccc}
          32 & 3 & 0 & 0 & 0 & 0 & 9 \\
          27 & 26 & 52 & 0 & 0 & 0 & 62 \\
          0 & 55 & 63 & 39 & 0 & 0 & 35 \\
          0 & 0 & 99 & 12 & 24 & 0 & 71 \\
          0 & 0 & 0 & 74 & 61 & 51 & 53 \\
          0 & 0 & 0 & 0 & 1 & 68 & 42 \\
          29 & 65 & 9 & 45 & 72 & 59 & 33
        \end{array}
                \right]\left[\begin{array}{c}
                               x_1 \\
                               x_2 \\
                               x_3 \\
                               x_4 \\
                               x_5 \\
                               x_6 \\
                               x_7
                             \end{array}\right]=\left[\begin{array}{c}
                                           90 \\
                                           24 \\
                                           43 \\
                                           97 \\
                                           51 \\
                                           52 \\
                                           56
                                         \end{array}\right]$\\
\\
i) By applying the \textbf{BTLE} algorithm in [5], it yields\\
\hspace*{2.7cm}$\mathbf{x_{BTLE}}=[3.8709, -2.2703, 3.1714, 1.8793, -1.0861, 2.6376, -3.0065]^T$.\\
ii) By applying the \textbf{SBTLE} algorithm, it yields\\
\hspace*{2.7cm}$\mathbf{x_{SBTLE}}=[3.8638, -2.2838, 3.1464, 1.9121, -1.0871, 2.6192, -2.9767]^T$.\\
iii) By applying MATLAB command $A\backslash b$, it yields\\
\hspace*{2.7cm}$\mathbf{x_{MATLAB}}=[3.8638, -2.2838, 3.1464, 1.9121, -1.0871, 2.6192, -2.9767]^T$.\\
iv) By applying \textbf{SMWBTLE} algorithm , it yields\\
\hspace*{2.7cm}$\mathbf{x_{SMW}}=[3.8638, -2.2838, 3.1464, 1.9121, -1.0871, 2.6192, -2.9767]^T$.\\
\\
\textbf{Example 3.2.} Let\\
\hspace*{2.5cm}$
\left[\begin{array}{cccccccccc}
        0 & 2 & 0 & 0 & 0 & 0 & 0 & 0 & 0 & 5 \\
        13 & 2 & 12 & 0 & 0 & 0 & 0 & 0 & 0 & 3 \\
        0 & 9 & 1 & 5 & 0 & 0 & 0 & 0 & 0 & 2 \\
        0 & 0 & 3 & 15 & 1 & 0 & 0 & 0 & 0 & 1 \\
        0 & 0 & 0 & 2 & 3 & 10 & 0 & 0 & 0 & 5 \\
        0 & 0 & 0 & 0 & 7 & 1 & 2 & 0 & 0 & 2 \\
        0 & 0 & 0 & 0 & 0 & -5 & 2 & 2 & 0 & 7 \\
        0 & 0 & 0 & 0 & 0 & 0 & 2 & 1 & 1 & 12 \\
        0 & 0 & 0 & 0 & 0 & 0 & 0 & 5 & 2 & 4 \\
        3 & 2 & 1 & 7 & 5 & -2 & 4 & 2 & 1 & 5
      \end{array}
                \right]\left[\begin{array}{c}
                               x_1 \\
                               x_2 \\
                               x_3 \\
                               x_4 \\
                               x_5 \\
                               x_6 \\
                               x_7\\
                               x_8\\
                               x_9\\
                               x_{10}
                             \end{array}\right]=\left[\begin{array}{c}
                                           7 \\
                                           30 \\
                                           17 \\
                                           20 \\
                                           30 \\
                                           12 \\
                                           6\\
                                           16\\
                                           11\\
                                           28
                                           \end{array}\right]$\\
\\
i) By applying the \textbf{BTLE} algorithm in [5], it breaks down\\
ii) By applying the \textbf{SBTLE} algorithm , it yields\\
$\mathbf{x_{SBTLE}}=[\frac{12067595}{8154227\,t+12067595 },{\frac {8516457\,
t+12067595}{8154227\,t+12067595}},{\frac {17567204\,t+12067595}{
8154227\,t+12067595}},{\frac {6642982\,t+12067595}{8154227\,t+12067595
}},{\frac {5142382\,t+12067595}{8154227\,t+12067595}},\\
\hspace*{1.7cm}{\frac {9396731
\,t+12067595}{8154227\,t+12067595}},{\frac {41265687\,t+24135190}
{2(8154227\,t+12067595)}},{\frac {7736309\,t+12067595}{8154227\,t+
12067595}},{\frac {14315844\,t+12067595}{8154227\,t+12067595}},{\frac
{5595816\,t+12067595}{8154227\,t+12067595}}]_{t=0}^T\\
\hspace*{1.24cm}=[1, 1, 1, 1, 1, 1, 1, 1, 1, 1]^T
$.\\
\\
\textbf{Example 3.3.} We consider the following general bordered linear system in order to demonstrate the efficiency of \textbf{SBTLE} algorithm and \textbf{SMWBTLE} algorithm. Let\\
\\
$$
\left[
  \begin{array}{ccccccc}
    2 & 3 & 0 & \cdots & \cdots & 0 & 4 \\
    1 & 2 & 3 & 0 & \ddots    & 0 & 4 \\
    0 & 1 & 2 & 3 & \ddots & \vdots     & \vdots \\
    \vdots & \ddots & \ddots & \ddots & \ddots  & 0 & 4\\
    \vdots & \ddots & \ddots & \ddots & \ddots  & \ddots & 4 \\
     0 &   \cdots & \cdots & 0 & 1 & 2 & 3 \\
    5 & 5 &  \cdots & \cdots & 5 & 1 & 2 \\
  \end{array}
\right]\left[
               \begin{array}{c}
                 x_1 \\
                 x_2 \\
                 x_3 \\
                 \vdots \\
                 x_{n-2} \\
                 x_{n-1} \\
                 x_n \\
               \end{array}
\right]=\left[\begin{array}{c}
                 9 \\
                 10 \\
                 10 \\
                 \vdots \\
                 10 \\
                 6 \\
                 5n-7 \\
               \end{array}\right]\\
$$
It can be verified that the exact solution is $x = [1, 1,\cdots,1]^T$. We used proposed algorimths, \textbf{SBTLE} and \textbf{SMWBTLE} Algorithms, and Gauss Elimination Algorithm to compute the solution, $\bar{x}$. Results are given in the next table in which $\varepsilon=||x-\bar{x}||_\infty$.\\

$$Table1.$$
$$
\begin{tabular}{|c|c|c|c|}
  \hline

  \multirow{2}{*}{n} & \multicolumn{3}{c|}{$\varepsilon=||x-\bar{x}||_\infty$ and CPU time} \\ \cline{2-4}
   & \textbf{SBTLE} Algorithm & \textbf{SMWBTLE} Algorithm & Gauss Elimination Algorithm \\ \hline
  500 & $3.4100\times 10^{-8}|\quad  0.0613$ & $1.4600\times 10^{-7}|\quad  0.1633$ & $4.4000\times 10^{-8}|\quad  0.0616$ \\ \hline
  000 & $6.9100\times 10^{-8}|\quad  0.1113$ & $2.9600\times 10^{-7}|\quad  0.3133$ & $8.9000\times 10^{-8}|\quad  0.1116$ \\ \hline
  5000 & $3.4910\times 10^{-7}|\quad  0.5113$ & $1.4960\times 10^{-6}|\quad 1.5133$ & $4.4900\times 10^{-7}|\quad  0.5116$ \\ \hline
  10000 & $6.9910\times 10^{-7}|\quad 1.0113$ & $2.9960\times 10^{-6}|\quad  3.0133$ & $8.9900\times 10^{-7}|\quad  1.0116$ \\
  \hline
 \end{tabular}$$

\section{CONCLUSIONS}

In this work new symbolic computational algorithms have been developed for computing the solution of a general bordered tridiagonal system. The algorithms are reliable, computationally efficient and remove the cases where the numeric algorithms are fail.

\end{document}